# A Fast Percolation-Dijkstra Routing Method for Mega-Constellation Backbone Network

Shenshen Luan, Member, IEEE, Luyuan Wang, Yepeng Liu, Ninghan Sun, Ran Zhang

***Abstract*—The real-time routing for satellite communication of the mega-constellations is being challenged due to the large-scale of network nodes, especially on devices with limited computation such as onboard embedded systems. In this paper, a fast routing method is proposed for mega-constellation backbone networks. Firstly, inspired by the regularity and sparse characteristics of mega-constellations, the 4-degree percolation theory is proposed to describe the node search process. Then, dynamic minimum search and mapping methods are used to narrow down the traversal range. The proposed method performs as well as the heap-optimized Dijkstra algorithm with less memory space and dynamic access. The experimental results show that the method proposed in this paper can significantly reduce routing computation time, especially on the onboard, edge-computing or other computation-limited devices.**

***Index Terms*—mega-constellation routing, 4-degree percolation, routing algorithm, satellite network, Dijkstra algorithm.**

## I. INTRODUCTION

In recent years, satellite Internet built on the low-Earth orbit(LEO) mega-constellations has developed rapidly[1], such as Starlink, Kuiper, Oneweb et.al. In the space segment, the satellite network is mainly composed of mega-constellation backbone networks, which contains a large number of satellite nodes. For example, there are 72x22 satellites in Starlink K1 layer and 34x34 in Kuiper K1 layer. When the scale of network nodes is extremely increasing, it usually leads to slower routing calculation time [2-3], which is particularly severe on embedded devices with limited computing resource such as onboard computers. Therefore, how to improve the efficiency of large-scale network routing calculation on the onboard computers has become one of the most important problems to be solved.

The Shortest-Distance Path (SPD) algorithm is one of the classic methods for network routing calculation and has been used in the mega-constellations network [4-6]. The Dijkstra algorithm adopts a greedy search strategy to traverse all nodes in the network for shortest path calculation and search. But the traversing operating without the consideration of the topology has led to its poor real-time performance in solving sparse-graph routing problem. Eramo V et.al.[7] proposed a heap-optimization method to improve the Dijkstra algorithm to reduce time complexity in sparse graphs. But the heap operation and complex data structures increase computational costs. Zhang et al. [8] proposed using Dijkstra to calculate the minimum hop routing in satellite and ground integrated networks in the architecture of large-scale network routing algorithms. For multi-layer networks, Ma et.al. [9] proposed using Dijkstra to obtain the optimal routing table for MEO satellites to achieve domain clustering of LEO satellites. Dijkstra has also been used to address advantageous switching sequences in graph-based user-satellite switching frameworks in LEO satellite networks [10]. Due to complex data structures and traversing operations, the above methods often cannot meet the real-time requirements of network communication on embedded platforms or systems. Chen et al. [11-12] proposed an analytical model-based shortest path routing algorithm considering the topology characteristics of mega-constellation networks, which effectively reduces routing computation time. Pan et al. [13] proposed an Orbit Prediction Shortest Path First (OPSPF) algorithm for elastic LEO satellite networks using the terrestrial routing protocol OSPF. Li et al. [14] described a time-varying topology for random transmission requirements and random packet generation/arrival, and proposed an effective Net-grid based Shortest Path Routing (NSPR) algorithm for large-scale satellite networks. However, due to the high complexity of the algorithm, it is still not suitable for the operation of onboard embedded devices.

In addition to large-scale nodes, the limited computing resource of the onboard embedded computer in satellite network is also the key considerations in designing routing algorithms. Inspired by the regularity and sparsity characteristics, a Percolation Dijkstra algorithm is proposed to accelerate the routing calculation of the mega-constellation backbone network, including four-degree percolation and dynamic min-search. Section II introduces the proposed method and its detailed procedures in this article. Section III illustrates the time complexity analysis of the proposed methods and comparation with other algorithms; The demonstration results on embedded platforms and discussion are listed in the section IV and the last section conclude all the method and contribution in this article.

Shenshen Luan is with Electronic and Information Engineering School, Beihang University, Beijing, 100191, China (e-mail: luanshenshen@buaa.edu.cn).

Luyuan Wang is with Department of Electronic Engineering and Information Science, University of Science and Technology of China, 230022, (email: wang_luyuan@sina.com).

Yepeng Liu, Ninghan Sun and Ran Zhang are with School of Information and Communication Engineering, Beijing University of Posts and Telecommunications (email: LiuYepeng2020@gmail.com, sunnh@bupt.edu.cn, zhangran@bupt.edu.cn.).

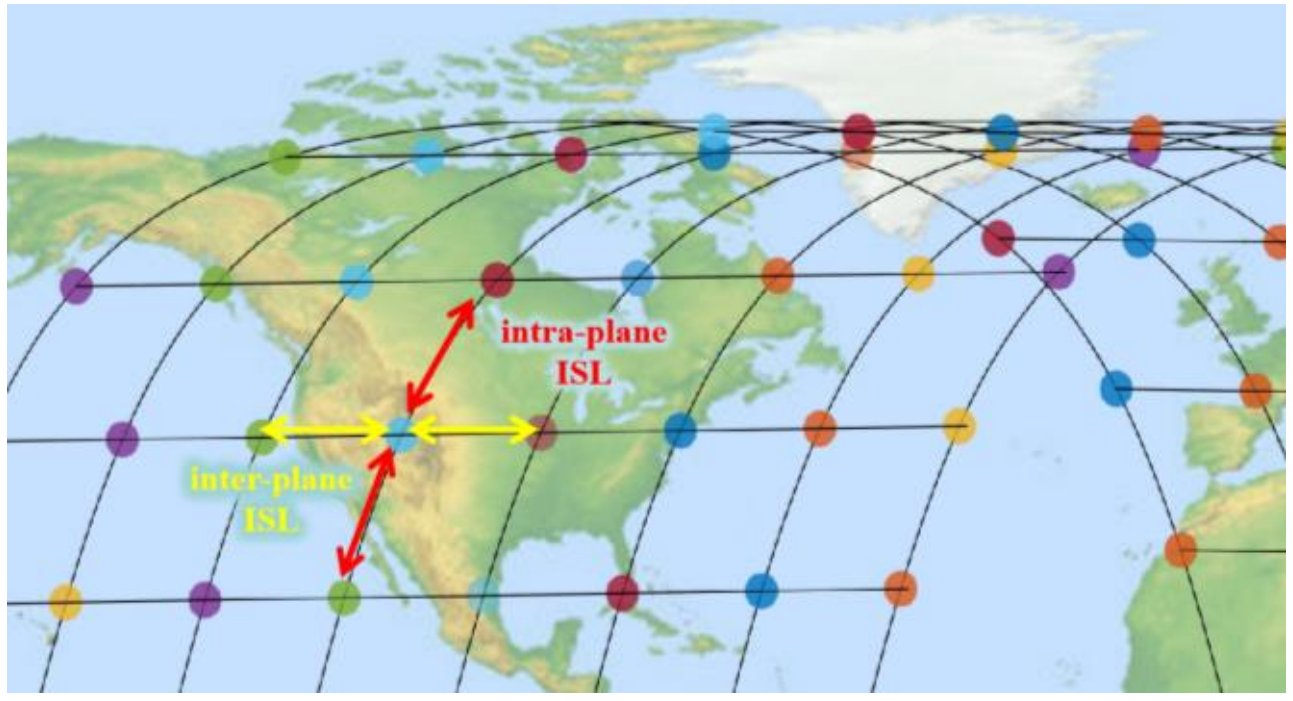

**Fig. 1.** The mesh-like topology of mega-constellation backbone network and the inter/intra plane ISL of the cross shaped unit..

## II. Methodology

### *A. Adjacent Matrix of Mega-Constellation*

The graph-based routing algorithm usually describes the network topology with nodes and edges, which is programmed using link lists or priority queues with C language. In the proposed method, an adjacency matrix is used to describe the network topology and serves as an input variable for programming implementation. For a mega-constellation composes of M orbits and N satellite in each orbit, the number of rows and columns of its adjacency matrix is M*N respectively. The element values of the adjacency matrix represent the cost between any two satellite nodes, where diagonal element 0 represents the satellite node itself, floating-point numbers from 0 to 1 represent link cost, and NULL represents the inability of the two satellites to connect.

Figure 2 shows an adjacent matrix of a 3*5 topology, which means 3 orbits and 5 satellites in each orbit. The cost here is set as 1 for convenience. As figure 2 shows, the element with gray background means the inter-plane ISL link cost. The element with blue background is the intra-plane ISL link cost and the orange-background elements illustrate the link cost between the first plane and the last plane ISL. It can be seen that the link cost distribute along the diagonal line of the matrix and there are 6 main link-cost lines in the matrix no matter how large the size of the matrix is. This is the sparse characteristics of the mega-constellation. Although the inter-plane link will be connected or disconnected during the movements of the satellites, the link cost in the adjacent matrix only changes its values instead of changing its positions, which is so called the regularity characteristics.

In addition, a percolation array, a visited array, and a distance array will be set up to record the percolation results, visited node index and the current shortest distance respectively, which is similar with Dijkstra algorithm. The initial value of the percolation array is set to null. The percolation array is length-varying with the input/output of the visited nodes during the iteration of the percolation, which can help reduce the traversing times of the minimum search.

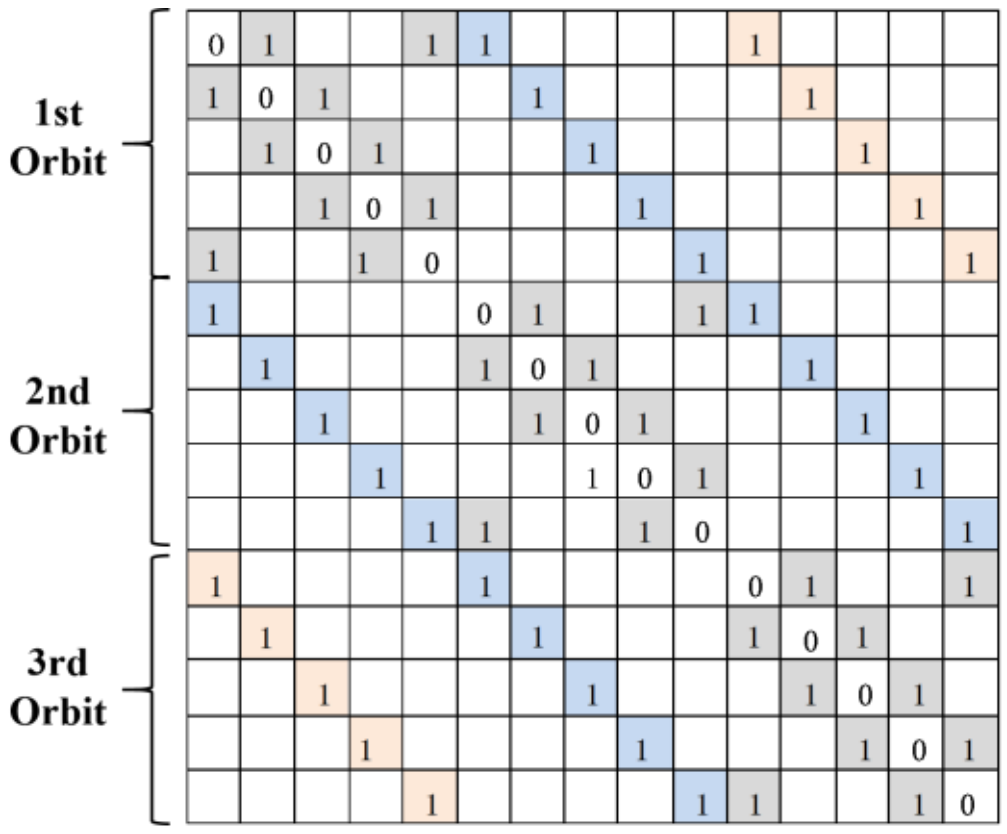

**Fig. 2.** The adjacent matrix of mega-constellation with link cost of different ISLs.

### *B. 4-degree Percolation*

The Dijkstra-based algorithm usually needs to traverse all remaining nodes when searching for neighboring nodes, resulting in the computation equal to the number of nodes N. According to the 4-degree sparse feature of the backbone network, the searching should percolate along a regular path instead of traversing all the nodes. In the adjacency matrix, searching for the connected nodes of the i-th satellite only requires calculating and manipulating the values of the matrix element A[i, j], where j belongs to [i-1, i+1, i+N, i-N]. A[i, j] is the adjacent matrix and N is the satellite number of each orbit. If $distance(\mathrm{i}) + \mathrm{w}(\mathrm{i},\mathrm{j}) < distance(j)$, the satellite j is the next shortest-distance node. Add the index j to the percolation array and update the distance to the corresponding position in the distance array.

Figure 3 shows an example of finding the shortest-distance path from node 1 to node 9 in a mesh network. Every column of the network topology is in the same orbit plane of the mega-constellation. When the starting node 1 searching for the next shortest-distance node, it just needs to check the link costs with the neighbor 4 nodes, other than all the 25 nodes. It works like the percolation rather than the flooding of water on the network topology. According to the shortest-distance principle, node 2 is the next staring node. It can be seen that with this strategy, the shortest-distance path can be found after 4 times percolation and the shortest-distance is 1-2-5-8-9.

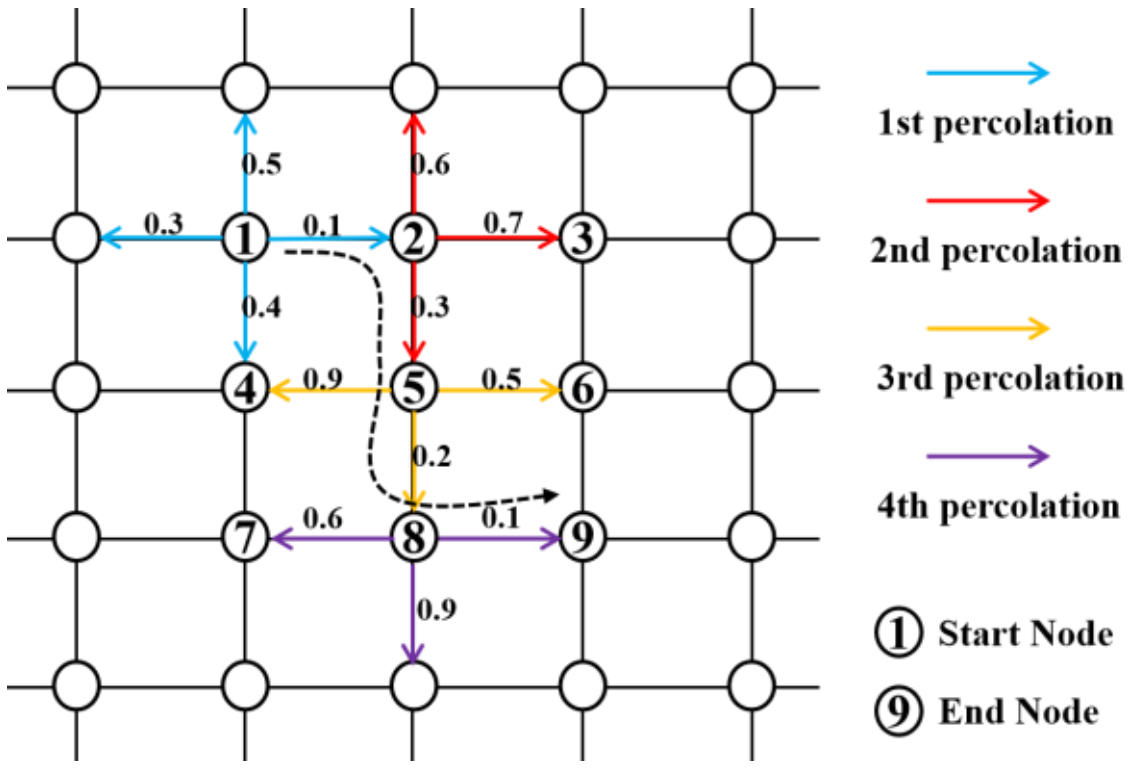

**Fig. 3.** Example of finding the shortest-distance path on the mesh topology using 4-degree percolation .

### *C. Dynamic Min-Search*

After updating the percolation array, the shortest distance node should be searched and set as the next node of 4-degree percolation. As shown in Figure 3, traverse all the elements in the percolation array and search for the corresponding minimum value in the distance array. Then record the min-value's node index α and set the visited array value of the node α to 1. Finally, set the α as the next node to perform 4-degree percolation and remove α from the percolation array. As iterations continue, the number of elements in the percolation array dynamically changes. Figure 4 shows the flowchart of the proposed method.

**Algorithm 1** Dynamic Min-Research

**Input:** Percolation_array
**Output:** Percolation_array, Next_node, Visited_array

```
SourceNode = 0;
Min = INF;
FOR I = 1 : length(Percolation_array)
    Percolation_node = Percolation_array(i);
    IF Min > distance_array(Percolation_node)
        Min = distance_array(Percolation_node);
        SourceNode = Percolation_node;
    END
END
IF SourceNode != 0
    Visited_array(SourceNode) = 1;
    Next_node = SourceNode;
    Remove Percolation_mode from
Percolation_array
END
```

## III. COMPLEXITY ANALYSIS

To verify the effectiveness of the proposed method theoretically, we analyzed its time complexity and compared it with the Dijkstra algorithm and give the efficiency express compared with Dijkstra.

In the 4-degree percolation method, due to the limitation of the maximum node number, the calculation times of each node is reduced from N to 4. In the dynamic minimum search method, due to the dynamic number of elements in the percolation array, a quantitative description of the calculation times X cannot be determined. But X can be estimated averagely using the Monte Carlo algorithm. Based on the experimental results in the next section, it can be determined that the value of X is approximately N/7.5 after 1000 times experiments. Therefore, the total computation times of the proposed method is (4+X)* N. By comparing the computation times of the Dijkstra algorithm with 2*N2, the algorithm's computation complexity has been effectively improved. The efficiency of the proposed method compared with Dijkstra can be expressed as following. It can be concluded that the maximum efficiency can reach 15 when the node number N tends to positive infinity theoretically.

$$\eta = \frac{(4+X)*N}{2*N^2} = \frac{4+X}{2*N} \tag{1}$$

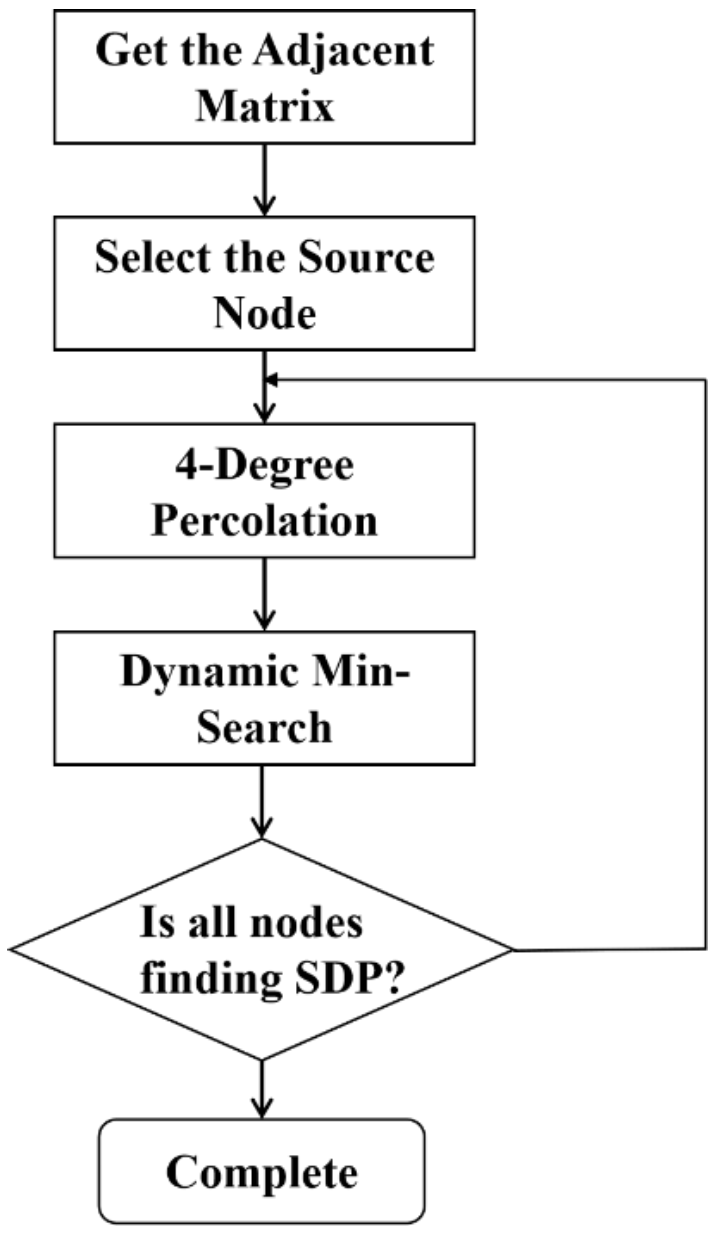

**Fig. 4.** The flowchart of the proposed method.

## IV. EXPERIMENTS

The topology of the mega-constellation backbone network is similar with that of the mesh-grid, the network scale can be expanded by increasing the number of constellation orbits. In this paper, the constellation scale is expanded by increasing orbit number from 3 to 36 with 18 satellites per orbit. The proposed algorithm, Dijkstra algorithm, and heap optimized algorithm was conducted to verified the effectiveness of the proposed method. The verification is demonstrated on the Icore4T embedded edge computing platform, with 11th Gen Intel(R) Core(TM) i5-11300H @ 3.10GHz and 16GB RAM. All the three algorithms are implemented by C/C++ language. The results are shown in Figure 5. It can be seen that the proposed algorithm has greatly improved with the constellation scale increasing compared to the Dijkstra algorithm. The comparison between the proposed method and heap-optimized Dijkstra algorithm is also shown as the ratio of their calculation time as shown in Figure 6. It can be seen that when the network scale is less than 334, the proposed method outperforms much better. When the network scale is increasing over 334, the calculation time of the proposed method is almost a little more than the other and stable with the network scale increases.

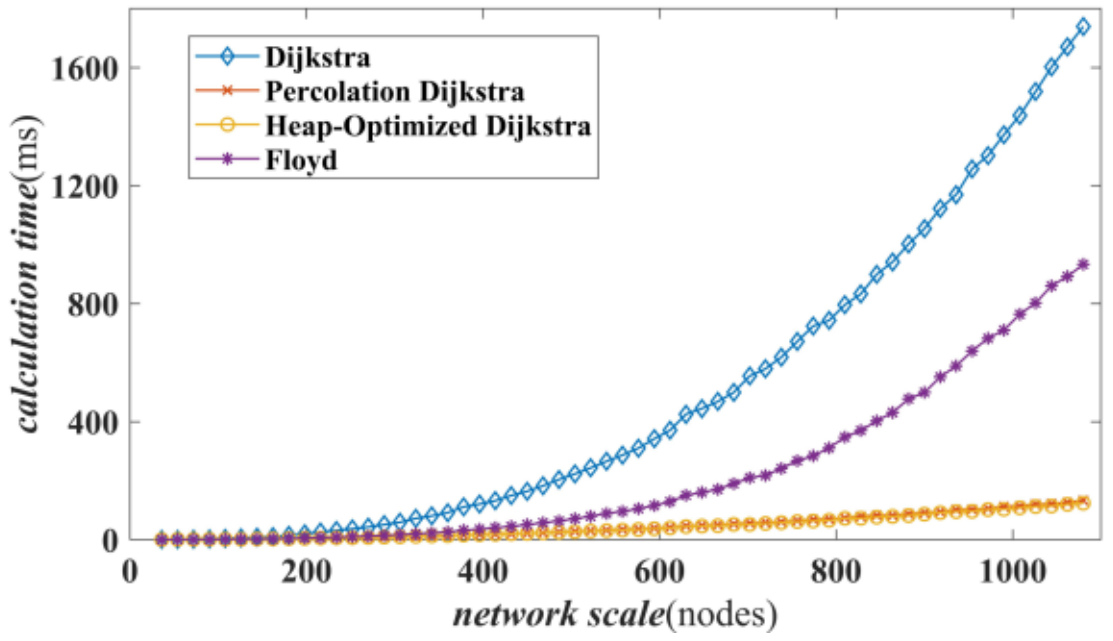

**Fig. 5.** Calculation time of algorithms with increasing network scale by 18 satellites per orbit.

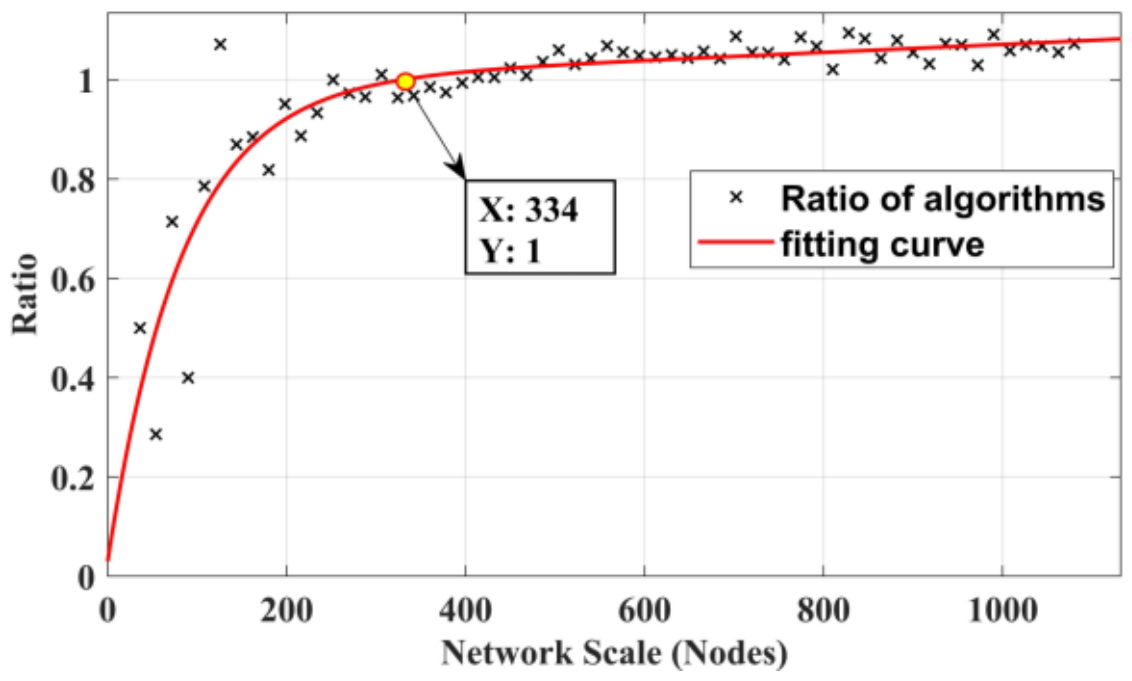

**Fig. 6.** The ratio of the calculation time between the proposed method and the heap-optimized Dijkstra algorithm.

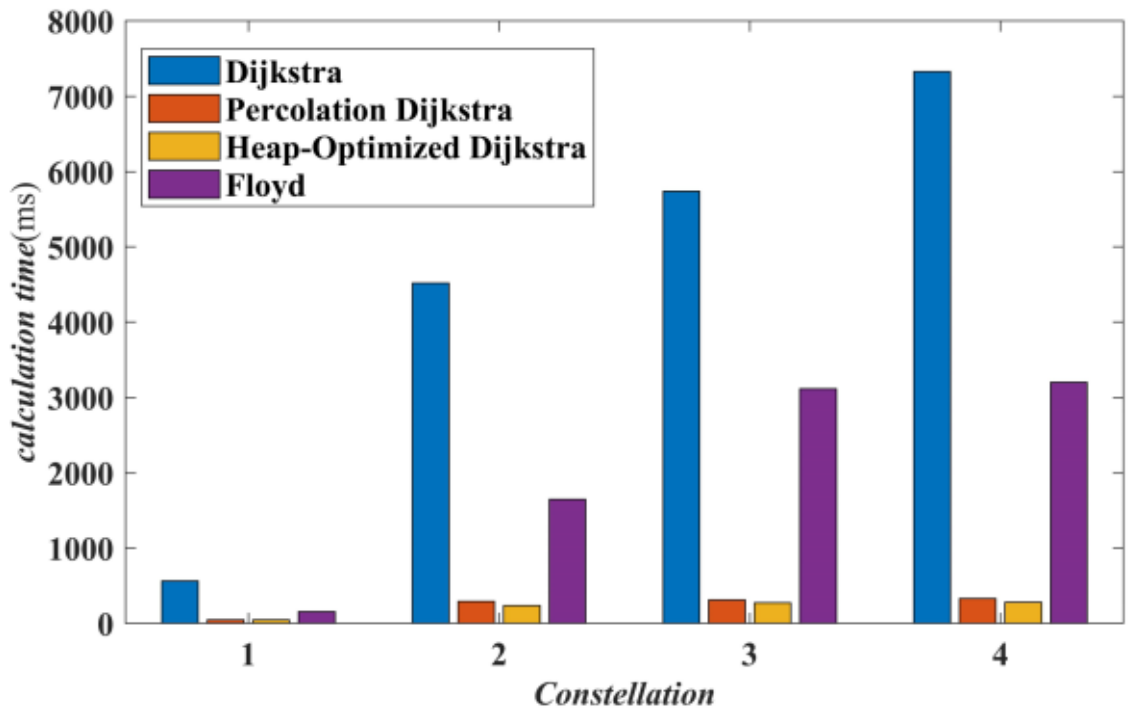

**Fig. 7.** The calculation time of typical constellation topology. 1 for Oneweb, 2 for Kuiper, 3 for Starlink-a layer, 4 for Starlink-b layer.

We also conducted experiments on typical constellation topologies, such as Oneweb (18x36), Kuiper (34x34), Starlink-a (72x22), and Starlink-b (24x66). It can be seen from the Figure 7 that the proposed method can get the same performance with the heap-optimized Dijkstra algorithm while avoiding a lot of memory access and space with heap or prior queue data structure. In practice, there is usually no such a large network scale reaching nearly 1000 nodes for communication networks. The proposed network can get the best performance within 334 nodes as shown in Figure 6.

To verify the proposed method on the embedded devices, experiments on the onboard computing platform with SPARC v8 CPU @133MHz with no operation system environment is also conducted. The results shown in the Figure 8 illustrate that the proposed method has a good consistence with the analysis in section III and can promote the real-time performance greatly without introduce complex data structure.

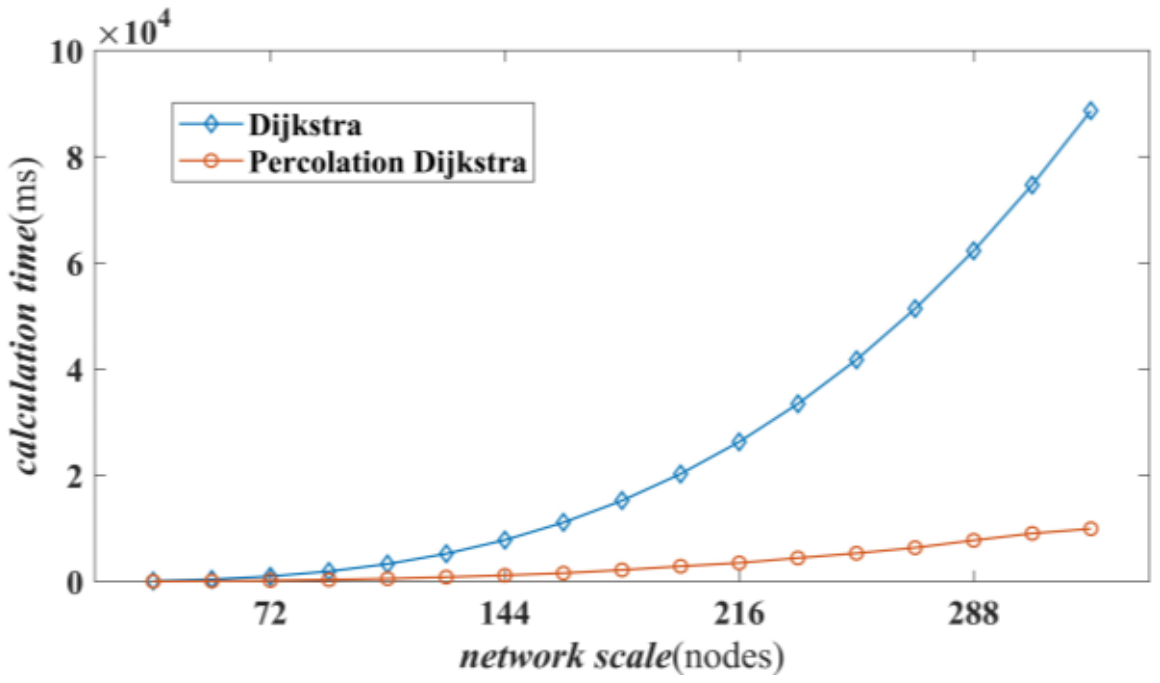

**Fig. 7.** The calculation time on onboard device.

## V. Conclusion

In this article, a fast routing algorithm is proposed to accelerate the routing calculation of the large-scale backbone network of mega-constellation. Firstly, based on the sparsity and regularity characteristics of the mega-constellation, a 4-degree percolation algorithm was designed. Secondly, a dynamic minimum search algorithm was proposed to limit the range of traversal calculations. The experimental results show that compared to conventional algorithms, the method proposed in this paper can significantly reduce routing computation time and is suitable for satellite computers with limited computing resources.

## References

[1] Xiong T, Zhang R, Liu J, et al. A blockchain-based and privacy-preserved authentication scheme for inter-constellation collaboration in Space-Ground Integrated Networks[J]. Computer Networks, 2022, 206: 108793.

[2] Liu J, Zhang X, Zhang R, et al. Reliable and low-overhead clustering in LEO small satellite networks[J]. IEEE Internet of Things Journal, 2021, 9(16): 14844-14856.

[3] Zhou Y, Zhang R, Liu J, et al. A Hierarchical Digital Twin Network for Satellite Communication Networks[J]. IEEE Communications Magazine, 2023.

[4] A. U. Chaudhry and H. Yanikomeroglu, "When to Crossover From Earth to Space for Lower Latency Data Communications?," in IEEE Transactions on Aerospace and Electronic Systems, vol. 58, no. 5, pp. 3962-3978, Oct. 2022, doi: 10.1109/TAES.2022.3156087.

[5] Z. Lin et al., "Systematic Utilization Analysis of Mega-Constellation Networks," 2022 International Wireless Communications and Mobile Computing (IWCMC), Dubrovnik, Croatia, 2022, pp. 1317-1322, doi: 10.1109/IWCMC55113.2022.9824193..

[6] X. Qin, T. Ma, Z. Tang, X. Zhang, H. Zhou and L. Zhao, "Service-Aware Resource Orchestration in Ultra-Dense LEO Satellite-Terrestrial Integrated 6G: A Service Function Chain Approach," in IEEE Transactions on Wireless Communications, vol. 22, no. 9, pp. 6003-6017, Sept. 2023, doi: 10.1109/TWC.2023.3239080.

[7] Eramo V , Listanti M , Caione N ,et al.Optimization in the Shortest Path First Computation for the Routing Software GNU Zebra[J].IEICE Transactions on Communications, 2005.

[8] S. Zhang and K. L. Yeung, "Scalable routing in low-earth orbit satellite constellations: Architecture and algorithms", Comput. Commun., vol. 188, pp. 26-38, Apr. 2022.

[9] T. Ma, B. Qian, X. Qin, X. Liu, H. Zhou and L. Zhao, "Satellite-Terrestrial Integrated 6G: An Ultra-Dense LEO Networking Management Architecture," in IEEE Wireless Communications, vol. 31, no. 1, pp. 62-69, February 2024, doi: 10.1109/MWC.011.2200198.

[10] M. Hozayen, T. Darwish, G. K. Kurt and H. Yanikomeroglu, "A Graph-Based Customizable Handover Framework for LEO Satellite Networks," 2022 IEEE Globecom Workshops (GC Wkshps), Rio de Janeiro, Brazil, 2022, pp. 868-873, doi: 10.1109/GCWkshps56602.2022.10008514.

[11] Q. Chen, L. Yang, Y. Zhao, Y. Wang, H. Zhou and X. Chen, "Shortest Path in LEO Satellite Constellation Networks: An Explicit Analytic Approach," in IEEE Journal on Selected Areas in Communications, doi: 10.1109/JSAC.2024.3365873..

[12] Q. Chen, G. Giambene, L. Yang, C. Fan and X. Chen, "Analysis of Inter-Satellite Link Paths for LEO Mega-Constellation Networks," in IEEE Transactions on Vehicular Technology, vol. 70, no. 3, pp. 2743-2755, March 2021, doi: 10.1109/TVT.2021.3058126.

[13] T. Pan, T. Huang, X. Li, Y. Chen, W. Xue, and Y. Liu, "OPSPF: Orbit Prediction Shortest Path First Routing for Resilient LEO Satellite Networks," in ICC 2019 - 2019 IEEE International Conference on Communications (ICC), 2019, pp. 1–6.

[14] J. Li, H. Lu, K. Xue, and Y. Zhang, "Temporal Netgrid ModelBased Dynamic Routing in Large-Scale Small Satellite Networks," IEEE Transactions on Vehicular Technology, vol. 68, no. 6, pp. 6009–6021, Jun. 2019.